\documentclass[10pt,a4paper,twocolumn]{revtex4}
\usepackage[latin1]{inputenc}
\usepackage{amsmath}
\usepackage{amsfonts}
\usepackage{amssymb}
\usepackage{graphicx}

\begin{document}
\title{Generalized Floquet theory for open quantum systems}
\author{C. M. Dai, Hong Li, W. Wang, and X. X. Yi\footnote{yixx@nenu.edu.cn}}
\affiliation{Center for Quantum Sciences and School of Physics,
	Northeast Normal University, Changchun 130024, China}

\date{\today}

\begin{abstract}
For a periodically driven  open quantum system, the Floquet theorem states that the time evolution operator $\Lambda(t,0)$ of the system can be factorized  as $\Lambda(t,0)=\mathcal{D}(t)e^{\mathcal{L}_{eff}t}$ with micro-motion operator $\mathcal{D}(t)$ possessing  the same period  as the external driving, and time-independent operator  $\mathcal{L}_{eff}$. In this work, we extend this theorem to   open systems that follow  a modulated periodic evolution, in which the fast part is periodic while the slow part breaks the periodicity. We derive a factorization for the time evolution operator that separates the long time dynamics and the micro-motion for the open quantum system. High-frequency expansions for the effective evolution operator  control the long time dynamics, and the micro-motion operator is also given and discussed. It may find applications in quantum engineering with open systems following modulated periodic evolution.
\end{abstract}

\maketitle
\section{Introduction}	
Floquet theory has very long history. It can be dated back to 1880s in mathematics due to Gaston Floquet who gave  a canonical form of solution to   periodic linear differential equations. The application of Floquet theory in physics ranges from classical \cite{gammaitoni1998} to quantum systems \cite{shirley1965}, covering a variety of time-dependent dynamics. In  recent years, the concept of Floquet engineering has attracted much attention since in periodically driven systems exotic phenomena can emerge that are absent in their undriven counterparts. The Floquet engineering is based on the fact that when a quantum system subject to periodically driving fields, the time evolution is governed by a time-independent effective Hamilton apart from a micro-motion described by the time-periodic unitary operator \cite{bukov2015,eckardt2015}. This concept provides us with  a versatile tool to manipulate  quantum systems and has been employed successfully in experiments, such as the control of  superfluid-to-Mott-insulator transition \cite{zenesini2009,eckardt2005}, the realization of artificial magnetic fields and topological band structures \cite{struck2011,bermudez2011,Aidelsburger2011,struck2012,hauke2012,struck2013,aidelsburger2013,miyake2013,atala2014,aidelsburger2015} as well as the modulation of  spin-orbit couplings \cite{jim2015}, to mention a few of them.

The  Floquet theory has found its application not only in closed quantum system where the dynamics is governed by unitary  evolution, but also  in open quantum systems \cite{haddadfarshi2015,dai2016} undergoing  non-unitary evolution. Formally, an open system can be described as follows, when a quantum system is coupled to an environment, the evolution of the whole system(system plus bath) governed by the total  Hamiltonian $H_{total}(t)$ is unitary. We can get the exact system state  by tracing over the bath degrees of freedom \cite{breuer2002},
\begin{equation}\label{utotal}
\rho(t)=Tr_{bath}[U_{total}(t)\rho_{sb}(0)U^{\dagger}_{total}(t)],
\end{equation}
where $\rho(t)$ is the reduced density matrix for the system, $\rho_{sb}(0)=\rho(0)\otimes\rho_{b}(0)$ is the uncorrelated initial system-bath state, and $U_{total}(t)=\mathcal{T}exp[-i\int_{0}^{t}H_{total}(t')dt']$ ($\mathcal{T}$ denotes time-ordering here and hereafter). It is possible to cast Eq.(\ref{utotal}) into the convolutionless form by certain approximations \cite{breuer2002},
\begin{equation}\label{lrho}
\partial_{t}\rho=\mathcal{L}(t)\rho(t).
\end{equation}
An  example is the time-dependent Markovian process  governed by the generator $\mathcal{L}(t)$ in the Lindblad form \cite{lindblad1976}
\begin{equation}\label{lindblad}
\begin{aligned}
\partial_{t}\rho=&-i[H(t),\rho]+\sum_{\alpha}[V_{\alpha}(t)\rho V^{\dagger}_{\alpha}(t)\\
&-\frac{1}{2}V^{\dagger}_{\alpha}(t)V_{\alpha}(t)\rho-\frac{1}{2}\rho V^{\dagger}_{\alpha}(t)V_{\alpha}(t)],
\end{aligned}
\end{equation}
where $V_{\alpha}(t)$ ($\alpha=1,2,3,...$)are time-dependent operators determined by the system-bath interaction, and $H(t)$ is the time-dependent effective Hamiltonian of the open system. It has been demonstrated that the dynamics given by Eq.(\ref{lrho}) can always be embedded in a time-dependent Markovian dynamics on an appropriate extended state space \cite{breuer2004}. A large class of non-Markovian quantum processes in open systems can also be described by Eq.(\ref{lrho}).

Eq.(\ref{lrho}) yields a two-parameter map $\Lambda(t,s)$ defined by chronological time-ordering operator $\mathcal{T}$
\begin{equation}\label{dlambda}
\Lambda(t,s)=\mathcal{T}exp[\int_{s}^{t}\mathcal{L}(\tau)d\tau], t\geq s\geq 0,
\end{equation}
and it satisfies
\begin{equation}\label{flambda}
\Lambda(t,s)\Lambda(s,t')=\Lambda(t,t'), t\geq s\geq t'.
\end{equation}
In terms of these maps, the solution to the master equation Eq.(\ref{lrho}) can be written as $\rho(t)=\Lambda(t,0)\rho(0)$\cite{davies1978}. This means $\Lambda(t,s)$ propagates the density matrix at time $s$ to the density matrix at time $t$.

When the generator $\mathcal{L}(t)$ possesses discrete time translation symmetry,  namely $\mathcal{L}(t+T)=\mathcal{L}(t)$, here $T$ is the period. According to the Floquet theorem \cite{dai2016,boite2017,yudin2016}, $\Lambda(t,s)$ can be decomposed into two parts, one can be given by an effective time-independent generator that controls the long time evolution and another is periodic in time describing the periodic micro-motion of the driven system (called the micro-motion operator) \cite{dai2016, yudin2016}.

Physically, such periodic time dependence of generator $\mathcal{L}(t)$ can be realized, for example, by coupling a static system to periodic driving field \cite{boite2016,kamleitner2011} or via periodic modulation of the coupling strength between different parts of the system \cite{wang2016}. \textbf{In practical situations, however, the time periodic dynamics  may be changed in different ways. For instance, a time periodic driving might  be turned on at some instance of time  and the amplitude of the driving needs a time to  ramp up to a certain value. In this case, the dynamics of the system is not perfectly periodic.  It has been demonstrated experimentally that different ramping protocols can influence the Floquet state population \cite{desbuquois2017}. The other example is that, consider  atoms driven by laser pulses,  there may have chirp in the pulses, leading to frequency change in the pulses. This again breaks the periodicity of the dynamics.}

\textbf{In this manuscript, we consider a generalized Floquet formulism to handle the aforementioned problems, where the periodicity of generator $\mathcal{L}(t)$ is disturbed slightly  via other (slow-varying) time-dependent parameters,  and the frequency can be chirped and changes slowly. In the generalized formalism, we show that the propagator $\Lambda(t,s)$ can also be factorized into two parts, one is the long time evolution part given by an effective generator with slow time dependence, and another is micro-motion part with additional slowly changing terms.}

The remainder of this manuscript is organized as follows. In Sec. \rm{II}, we present a generalized Floquet formalism that separates the long time evolution and micro-motion. In Sec. \rm{III}, we calculate the effective generator and micro-motion operator by high frequency expansions. In Sec. \rm{IV}, we demonstrate our results with two examples. Conclusions and discussions are presented in Sec. \rm{V}.

\section{Formulism}
We start with the dynamic equation Eq.(\ref{lrho}) for  the density matrix. In our case, $\mathcal{L}(t)=\tilde{\mathcal{L}}[\theta+\omega t,\mathbf{p}(t)]$ and $\tilde{\mathcal{L}}$ is periodical  with period $2\pi$ with respect to the first argument, namely,
$\tilde{\mathcal{L}}[\theta+\omega t,\mathbf{p}(t)]=\tilde{\mathcal{L}}[\theta+\omega t+2n\pi,\mathbf{p}(t)]$ with integer $n$. Where the periodic time dependence of generator $\mathcal{L}(t)$ is introduced through   $\theta\rightarrow\theta+\omega t$, and $\mathbf{p}(t)$ represents a set of  time-dependent parameters that disturb the periodicity of $\mathcal{L}(t)$. In Sec. \rm{IV}, we will present two examples to show how  $\theta+\omega t$  and $\mathbf{p}(t)$ enter the dynamics of open systems. In the situation of chirped frequency, $\omega$   depends on time. Here we define $\omega_{eff}=\partial_{t}(\omega t)$ (called effective instantaneous frequency) that we will use later.

The formal solution of Eq.(\ref{lrho}) can be given by the propagator in Eq.(\ref{dlambda})
\begin{equation}\label{auxiliary equation}
\tilde{\rho}(t,\theta)\equiv\mathcal{T}exp[\int_{0}^{t}\tilde{\mathcal{L}}[\omega \tau+\theta,\mathbf{p}(\tau)]d\tau]\rho(0),
\end{equation}
or in the differential form with initial condition,
\begin{equation}\label{dauxiliary equation}
\partial_{t}\tilde{\rho}(t,\theta)=\tilde{\mathcal{L}}[\omega t+\theta,\mathbf{p}(t)]\tilde{\rho}(t,\theta), \tilde{\rho}(0,\theta)=\rho(0).
\end{equation}
We can think the Eq.(\ref{dauxiliary equation}) as a family of equations parameterized by initial phase $\theta$, the corresponding propagators $\Lambda_{\theta}(t,s)=\mathcal{T}exp[\int_{s}^{t}\tilde{\mathcal{L}}[\omega \tau+\theta,\mathbf{p}(\tau)]d\tau]$ are also dependent on parameter $\theta$.

We extend the physical Hilbert space $\mathcal{H}$ to Floquet space $\mathcal{H}\bigotimes\mathcal{F}$ \cite{peskin1993,fleischer2005,eckardt2015,sambe1973,goldman2014,novicenko2017,howland1974,holthaus1989,eckardt2008}. Where $\mathcal{F}$ is the space of square-integrable functions on the circle of length $2\pi$  with  scalar product defined by $\langle\xi_1|\xi_2\rangle=\int_{0}^{2\pi}\xi_{1}^{*}(\theta)\xi_{2}(\theta)d\theta/2\pi$. The space $\mathcal{F}$ can be spanned  by the orthonormal basis $\{e^{in\theta}\}$ with $n\in\mathbb{Z}$ (all integers).

On the Floquet space $\mathcal{H}\bigotimes\mathcal{F}$, we can define a Floquet generator and that the evolution it generates is essentially equivalent with Eq.(\ref{auxiliary equation}). The Floquet generator is defined as,
\begin{equation}\label{lf}
\tilde{\mathcal{L}}_{\mathcal{F}}(t)=-\omega_{eff} \frac{\partial}{\partial \theta}+\tilde{\mathcal{L}}[\theta, \mathbf{p}(t)].
\end{equation}
the corresponding propagator satisfies equation,
\begin{equation}
\partial_{t}\Lambda_{\mathcal{F}}(t,s)=\tilde{\mathcal{L}}_{\mathcal{F}}(t)\Lambda_{\mathcal{F}}(t,s),\Lambda_{\mathcal{F}}(s,s)=\mathbf{1}.
\end{equation}
The relation between $\Lambda_{\theta}(t,s)$ and $\Lambda_{\mathcal{F}}(t,s)$ is given by
\begin{equation}\label{rl}
\Lambda_{\theta}(t,s)=\mathcal{S}(\omega t)\Lambda_{\mathcal{F}}(t,s)\mathcal{S}(-\omega s),
\end{equation}
where
\begin{equation}
\mathcal{S}(\omega t)=e^{\omega t\partial / \partial \theta},
\end{equation}
is the shift operator respect to $\theta$. This relation can be easily  verified by the definition and notice that $\partial_{t}\mathcal{S}(\omega t)=\mathcal{S}(\omega t)\omega_{eff}\frac{\partial}{\partial\theta}$, $\mathcal{S}(-\omega t)\tilde{\mathcal{L}}[\omega t+\theta,\mathbf{p}(t)]\mathcal{S}(\omega t)=\tilde{\mathcal{L}}[\theta, \mathbf{p}(t)]$.

We can see from the relation Eq.(\ref{rl}) that the density matrix with initial condition $\tilde{\rho}(0,\theta)=\rho(0)$ propagates by generator $\tilde{\mathcal{L}}_{\mathcal{F}}(t)$ is equivalent to by $\tilde{\mathcal{L}}[\omega t+\theta,\mathbf{p}(t)]$ up to a shift transformation. More specific, we define $\tilde{\rho}_{\mathcal{F}}(t,\theta)$ by equation,
\begin{equation}\label{ls equation}
\partial_{t}\tilde{\rho}_{\mathcal{F}}(t,\theta)=\tilde{\mathcal{L}}_{\mathcal{F}}(t)\tilde{\rho}_{\mathcal{F}}(t,\theta),\tilde{\rho}_{\mathcal{F}}(0,\theta)=\rho(0),
\end{equation}
and we have
\begin{equation}\label{sr}
\tilde{\rho}(t,\theta)=\mathcal{S}(\omega t)\tilde{\rho}_{\mathcal{F}}(t,\theta).
\end{equation}

By this transformation, we can transfer the dynamics to a frame that  is independent of $\omega t$. Namely, the periodic time dependence introduced by $\omega t$ can be eliminated by the transformation (this elimination holds even $\omega$ is time dependent). Ideally, for fixed $\omega_{eff}$ and $\mathbf{p}(t)$,  $\tilde{\mathcal{L}}_{\mathcal{F}}$ is time independent. The slow variation of $\omega_{eff}$ and $\mathbf{p}(t)$ will introduce a slow time dependence to the Floquet generator.

To process, we expand $\tilde{\rho}_{\mathcal{F}}(t,\theta)$ by  basis $\{e^{in\theta}\}$ in the Floquet space $\mathcal{H}\bigotimes\mathcal{F}$,
\begin{equation}\label{rf}
\tilde{\rho}_{\mathcal{F}}(t,\theta)=
\sum_{n=-\infty}^{\infty}\tilde{\rho}_{\mathcal{F}}^{(n)}(t)e^{in\theta}.
\end{equation}
We obtain a set of equations for the expansion coefficients $\tilde{\rho}_{\mathcal{F}}^{(n)}(t)$ in the physical space $\mathcal{H}$.
\begin{equation}\label{fourier equation}
\partial_{t}\tilde{\rho}_{\mathcal{F}}^{(n)}(t)
=\sum_{m=-\infty}^{\infty}\tilde{\mathcal{L}}_{\mathcal{F}}^{(n,m)}(t)\tilde{\rho}_{\mathcal{F}}^{(m)}(t).
\end{equation}
where $\tilde{\mathcal{L}}_{\mathcal{F}}^{(n,m)}(t)\equiv\int_{0}^{2\pi}e^{-in\theta}\tilde{\mathcal{L}}_{\mathcal{F}}(t)e^{im\theta}d\theta/2\pi=\tilde{\mathcal{L}}^{(n-m)}[\mathbf{p}(t)]-in\omega_{eff}\delta^{n}_{m}$ with $\tilde{\mathcal{L}}^{(n)}[\mathbf{p}(t)]\equiv\int_{0}^{2\pi}\tilde{\mathcal{L}}[\theta, \mathbf{p}(t)]e^{-in\theta}d\theta/2\pi$.

Define a vector by the coefficients $\tilde{\rho}_{\mathcal{F}}^{(n)}(t)$,
\begin{displaymath}
\tilde{\rho}_{\mathcal{F}}(t)=[\cdots, \tilde{\rho}_{\mathcal{F}}^{(-1)}(t), \tilde{\rho}_{\mathcal{F}}^{(0)}(t), \tilde{\rho}_{\mathcal{F}}^{(1)}(t), \cdots],
\end{displaymath}
we can write  Eq.(\ref{fourier equation}) as
\begin{equation}\label{fourier equationc}
\partial_{t}\tilde{\rho}_{\mathcal{F}}(t)=\tilde{\mathcal{L}}_{\mathcal{F}}(t)\tilde{\rho}_{\mathcal{F}}(t),
\end{equation}
here we use the same symbol $\tilde{\mathcal{L}}_{\mathcal{F}}(t)$ to represent the matrix form of $\tilde{\mathcal{L}}_{\mathcal{F}}(t)$ in Eq.(\ref{lf}) with the basis $\{e^{in\theta}\}$.

With the  shift matrix $\mathcal{R}_{l}^{(n,m)}=\delta_{m+l}^{n}$ and the number matrix $\mathcal{N}^{(n,m)}=n\delta_{m}^{n}$, $\tilde{\mathcal{L}}_{\mathcal{F}}(t)$ can be written in a more compact form,
\begin{equation}\label{lsc}
\tilde{\mathcal{L}}_{\mathcal{F}}(t)=\sum_{n=-\infty}^{\infty}
\tilde{\mathcal{L}}^{(n)}[\mathbf{p}(t)]\mathcal{R}_{n}-i\omega_{eff}\mathcal{N}.
\end{equation}
This means that $\tilde{\mathcal{L}}_{\mathcal{F}}(t)$ takes the following form
\begin{equation}\label{lscm}
\left[
\begin{matrix}
\ddots&\ddots&\ddots&\ddots&\ddots&\\
\ddots&\tilde{\mathcal{L}}^{(0)}+i\omega_{eff}&\tilde{\mathcal{L}}^{(-1)}&\tilde{\mathcal{L}}^{(-2)}&\ddots&\\
\ddots&\tilde{\mathcal{L}}^{(1)}&\tilde{\mathcal{L}}^{(0)}&\tilde{\mathcal{L}}^{(-1)}&\ddots&\\
\ddots&\tilde{\mathcal{L}}^{(2)}&\tilde{\mathcal{L}}^{(1)}&\tilde{\mathcal{L}}^{(0)}-i\omega_{eff}&\ddots&\\
\ddots&\ddots&\ddots&\ddots&\ddots&
\end{matrix}
\right].
\end{equation}

The matrix $\mathcal{R}_{n}$ and $\mathcal{N}$ in Eq.(\ref{lsc}) satisfy commutation relations $[\mathcal{R}_{l},\mathcal{R}_{k}]=0$, $[\mathcal{R}_{l},\mathcal{N}]=-l\mathcal{R}_{l}$ and $\mathcal{R}_{l}\mathcal{R}_{k}=\mathcal{R}_{l+k}$ with $l$ and $k$ arbitrary integers. These relations are useful when we derive a high frequency expansion in the Sec. \rm{III}.

We shall find a transformation $\mathcal{D}(t)$ that block diagonalize $\tilde{\mathcal{L}}_{\mathcal{F}}(t)$. This means with the transformation defined by $\mathcal{D}(t)$,
\begin{equation}\label{rhod}
\tilde{\rho}_{\mathcal{F}}(t)=\mathcal{D}(t)\tilde{\rho}_{\mathcal{D}}(t),
\end{equation}
the dynamic of $\tilde{\rho}_{\mathcal{D}}(t)$ is governed by a generator $\tilde{\mathcal{L}}_{\mathcal{D}}(t)$ of block diagonal form in the Floquet space $\mathcal{H}\bigotimes\mathcal{F}$,
\begin{equation}\label{diagonal equation}
\partial_{t}\tilde{\rho}_{\mathcal{D}}(t)=\tilde{\mathcal{L}}_{\mathcal{D}}(t)\tilde{\rho}_{\mathcal{D}}(t),
\end{equation}
with
\begin{equation}\label{ld}
\tilde{\mathcal{L}}_{\mathcal{D}}(t)=\mathcal{L}_{eff}(t)\mathcal{R}_{0}-i\omega_{eff}\mathcal{N},
\end{equation}
where $\mathcal{L}_{eff}(t)$ represents an effective generator in the physical space $\mathcal{H}$. Combining Eq.(\ref{fourier equationc}), Eq.(\ref{rhod}) and Eq.(\ref{diagonal equation}), $\tilde{\mathcal{L}}_{\mathcal{F}}(t)$ and $\tilde{\mathcal{L}}_{\mathcal{D}}(t)$  satisfy the following equation,
\begin{equation}
(\partial_{t}\mathcal{D}^{-1})\mathcal{D}+\mathcal{D}^{-1}\tilde{\mathcal{L}}_{\mathcal{F}}(t)\mathcal{D}=\tilde{\mathcal{L}}_{\mathcal{D}}(t).
\end{equation}
Thus $\tilde{\mathcal{L}}_{\mathcal{F}}(t)$ and $\mathcal{L}_{eff}(t)$ are connected through relation,
\begin{equation}\label{Leff equation}
(\partial_{t}\mathcal{D}^{-1})\mathcal{D}+\mathcal{D}^{-1}\tilde{\mathcal{L}}_{\mathcal{F}}(t)\mathcal{D}=\mathcal{L}_{eff}(t)\mathcal{R}_{0}-i\omega_{eff}\mathcal{N}.
\end{equation}
Due to the special structure of Floquet generator $\tilde{\mathcal{L}}_{\mathcal{F}}(t)$ Eq.(\ref{lsc}), if we find a transformation $\mathcal{D}(t)$ and the  corresponding  $\mathcal{L}_{eff}(t)$ satisfy Eq.(\ref{Leff equation}), $\mathcal{D}'(t)=\mathcal{R}_{l}^{-1}\mathcal{D}(t)\mathcal{R}_{l}$ is also a solution of Eq.(\ref{Leff equation}) with the same $\mathcal{L}_{eff}(t)$. It is sufficient to consider that $\mathcal{D}(t)$ is invariant under shift transformation $\mathcal{R}_{l}$ \cite{eckardt2015,novicenko2017}, this is the case when $\mathcal{D}(t)$ takes following form,
\begin{equation}\label{d}
\mathcal{D}(t)=\sum_{n=-\infty}^{\infty}\mathcal{D}^{(n)}(t)\mathcal{R}_{n}.
\end{equation}
When $\tilde{\mathcal{L}}_{\mathcal{F}}(t)$ is time independent, $\mathcal{D}(t)$ can be chosen to be time independent \cite{eckardt2015,haddadfarshi2015,dai2016,boite2017} and by solving Eq.(\ref{Leff equation}) we obtain the time independent effective generator $\mathcal{L}_{eff}$. This is exactly  the situation of conventional Floquet theory with the generator $\mathcal{L}(t)$ having perfect periodic time dependence. For the more general situation we consider here, the solution $\mathcal{D}(t)$ and $\mathcal{L}_{eff}(t)$ of Eq.(\ref{Leff equation}) may have explicit time dependence.

If $\mathcal{D}(t)$ has the form of Eq.(\ref{d}), its inverse  should also have the form $\mathcal{D}^{-1}(t)=\sum_{n=-\infty}^{\infty}{\mathcal{D}^{-1}}^{(n)}(t)\mathcal{R}_{n}$ (By the uniqueness of inverse and the equation $\mathcal{D}^{-1}\mathcal{D}=\mathbf{1}$ is invariant under shift transformation $\mathcal{R}_{l}$) where ${\mathcal{D}^{-1}}^{(n)}(t)$ is the expansion coefficients of ${\mathcal{D}^{-1}}(t)$ in the basis of shift matrix $\{\mathcal{R}_{n}\}$.

Consider the block diagonal form of $\tilde{\mathcal{L}}_{\mathcal{D}}(t)$, the formal solution for Eq.(\ref{diagonal equation}) is that
\begin{equation}
\tilde{\rho}_{\mathcal{D}}^{(n)}(t)=e^{-in\omega t}\Lambda_{eff}(t,0){\mathcal{D}^{-1}}^{(n)}(0)\rho(0),
\end{equation}
where $\Lambda_{eff}(t,0)=\mathcal{T}
exp[\int_{0}^{t}\mathcal{L}_{eff}(\tau)d\tau]$. What follows is
\begin{equation}
\begin{aligned}
\tilde{\rho}_{\mathcal{F}}^{(n)}(t)&=[\mathcal{D}(t)\tilde{\rho}_{\mathcal{D}}(t)]^{(n)}\\
&=\sum_{m=-\infty}^{\infty}\mathcal{D}^{(n-m)}(t)\tilde{\rho}_{\mathcal{D}}^{(m)}(t).
\end{aligned}
\end{equation}
Restoring  the basis $\{e^{in\theta}\}$ using Eq.(\ref{rf}) and performing  the shift $\mathcal{S}(\omega t)$ respect to $\theta$ using Eq.(\ref{sr}), we obtain
\begin{equation}\label{rhof}
\tilde{\rho}(t,\theta)=
\mathcal{D}(\theta+\omega t,t)\Lambda_{eff}(t,0)\mathcal{D}^{-1}(\theta,0)\rho(0),
\end{equation}
where $\mathcal{D}(\theta+\omega t,t)\equiv\sum_{n=-\infty}^{\infty}\mathcal{D}^{(n)}(t)e^{in(\theta+\omega t)}$ is the micro-motion operator depends on the initial phase $\theta$. Here the relation $\mathcal{D}^{-1}(\theta+\omega t,t)=\sum_{n=-\infty}^{\infty}{\mathcal{D}^{-1}}^{(n)}(t)e^{in(\theta+\omega t)}$ is used \cite{eckardt2015}.

The expression Eq.(\ref{rhof}) represents a generalized Floquet theory. For $\mathcal{D}$ on the Floquet space $\mathcal{H}\bigotimes\mathcal{F}$ without explicit time dependence, the coefficients $\mathcal{D}^{(n)}$ are time independent, the micro-motion operator $\mathcal{D}(\theta+\omega t,t)$ has periodic time dependence with period  $T=2\pi/\omega$. Generally, $\mathcal{D}(\theta+\omega t,t)$ will acquire additional time dependence due to the explicit time dependence of $\mathcal{D}^{(n)}(t)$.

Once we obtain the effective generator $\mathcal{L}_{eff}(t)$ and micro-motion operator $\mathcal{D}(\theta+\omega t,t)$, we can use Eq.(\ref{rhof}) to find the time evolution of density matrix. When $\mathcal{L}_{eff}(t)$ changes sufficiently slowly,  the long time evolution $\mathcal{T}exp[\int_{0}^{t}\mathcal{L}_{eff}(\tau)d\tau]$ can be treated  as a dynamics  governed  by the effective generator. Adiabatic approximation \cite{sarandy2005a,sarandy2005b,band1992} can be then applied straightforwardly.

Usually, $\mathcal{L}_{eff}(t)$ and $\mathcal{D}(\theta+\omega t,t)$ can not be determined analytically. But for sufficiently high instantaneous frequency $\omega_{eff}$, i.e.,  the operator norm of off-diagonal elements of $\tilde{\mathcal{L}}_{\mathcal{F}}(t)$ is much smaller than the instantaneous frequency $||\tilde{\mathcal{L}}^{(n)}[\mathbf{p}(t)]||\ll\omega_{eff}$ ($n\neq0$) and it changes little over one period $||\dot{\tilde{\mathcal{L}}}^{(n)}
[\mathbf{p}(t)]||\ll\omega_{eff}||\tilde{\mathcal{L}}^{(n)}[\mathbf{p}(t)]||$, both $\mathcal{L}_{eff}(t)$ and $\mathcal{D}(\theta+\omega t,t)$ can be represented as power series of inverse instantaneous frequency $\omega_{eff}^{-1}$ \cite{mikiami2016,blanes2009},  see the next section.

\section{High frequency expansion}
In this section we calculate $\mathcal{D}(\theta+\omega t,t)$ and $\mathcal{L}_{eff}(t)$ in the high frequency limit. Recall that matrix $\mathcal{D}(t)$ can be written as an exponential form $\mathcal{D}(t)=e^{\Omega(t)}$. Because $\mathcal{D}(t)$ has skew diagonal form, $\Omega(t)$ should have the same form $\Omega(t)=\sum_{n=-\infty}^{\infty}\Omega^{(n)}(t)\mathcal{R}_{n}$. Denote $\Omega(t)$ and $\mathcal{L}_{eff}(t)$ as a sum of different orders of $\omega_{eff}^{-1}$,
\begin{equation}\label{perturbation}
\begin{aligned}
\Omega(t)=&\sum_{n=1}^{\infty}\Omega_{(n)}(t),\\
\mathcal{L}_{eff}(t)=&\sum_{n=0}^{\infty}\mathcal{L}_{eff(n)}(t),
\end{aligned}
\end{equation}
where $n\textendash th$ terms $\Omega_{(n)}(t)=\sum_{m=-\infty}^{\infty}\Omega_{(n)}^{(m)}(t)\mathcal{R}_{m}$ and $\mathcal{L}_{eff(n)}(t)$ are of the order of $\omega_{eff}^{-n}$ (for simplicity we omit the argument $t$ hereafter). Take these into   Eq.(\ref{Leff equation}) and expand the left hand side of Eq.(\ref{Leff equation}) by identity  \cite{najfeld1995,blanes2009}
\begin{equation}\label{Leff right expansion}
\begin{aligned}
(\partial_{t}\mathcal{D}^{-1})\mathcal{D}&+\mathcal{D}^{-1}\tilde{\mathcal{L}}_{\mathcal{F}}\mathcal{D}=\\
\sum_{k=0}^{\infty}\frac{1}{(k+1)!}Ad_{-\Omega}^{k}[-\dot{\Omega}]&+\sum_{k=0}^{\infty}\frac{1}{k!}Ad_{-\Omega}^{k}[\tilde{\mathcal{L}}_{\mathcal{F}}],
\end{aligned}
\end{equation}
where $Ad_{-\Omega}^{k}[\mathcal{X}]$ means,
\begin{displaymath}
\overbrace{[-\Omega,\cdots[-\Omega,}^{k}\mathcal{X}]\cdots],
\end{displaymath}
we can derive expressions for the expansion in the following way.

To simplify the results,  we will set $\Omega^{(0)}_{(n)}(t)=0$ to find a special solution for the effective generator, because $\mathcal{L}_{eff}(t)$ is not uniquely identified by Eq.(\ref{Leff equation}) \cite{novicenko2017,mikiami2016}. Collecting  the same order terms in both sides of the Eq.(\ref{Leff equation}), we get a series of equations. The first three equations are
\begin{equation}\label{first three equation}
\begin{aligned}
&\mathcal{L}_{eff(0)}\mathcal{R}_{0}=\tilde{\mathcal{L}}+[\Omega_{(1)},i\omega_{eff}\mathcal{N}],\\
&\mathcal{L}_{eff(1)}\mathcal{R}_{0}=-\dot{\Omega}_{(1)}-[\Omega_{(1)},\tilde{\mathcal{L}}]+[\Omega_{(2)},i\omega_{eff}\mathcal{N}]\\
&-\frac{1}{2!}[\Omega_{(1)},[\Omega_{(1)},i\omega_{eff}\mathcal{N}]],\\
&\mathcal{L}_{eff(2)}\mathcal{R}_{0}=-\dot{\Omega}_{(2)}+\frac{1}{2!}[\Omega_{(1)},\dot{\Omega}_{(1)}]-[\Omega_{(2)},\tilde{\mathcal{L}}]\\
&+\frac{1}{2!}[\Omega_{(1)},[\Omega_{(1)},\tilde{\mathcal{L}}]]+[\Omega_{(3)},i\omega_{eff}\mathcal{N}]\\
&-\frac{1}{2!}[\Omega_{(1)},[\Omega_{(2)},i\omega_{eff}\mathcal{N}]]\\
&-\frac{1}{2!}[\Omega_{(2)},[\Omega_{(1)},i\omega_{eff}\mathcal{N}]]\\
&+\frac{1}{3!}[\Omega_{(1)},[\Omega_{(1)},[\Omega_{(1)},i\omega_{eff}\mathcal{N}]]],
\end{aligned}
\end{equation}
where $\tilde{\mathcal{L}}=\sum_{n=-\infty}^{\infty}
\tilde{\mathcal{L}}^{(n)}[\mathbf{p}(t)]\mathcal{R}_{n}$. Comparing  the coefficients in both sides  of Eq.(\ref{first three equation}) in terms of shift matrix, we can get $\Omega(t)$ and $\mathcal{L}_{eff}(t)$ up to the second order in $\omega_{eff}^{-1}$ and the higher order terms can be obtained in a similar way. The results are,
\begin{equation}\label{leff expansion}
\begin{aligned}
\mathcal{L}_{eff(0)}&=\tilde{\mathcal{L}}^{(0)},\\
\mathcal{L}_{eff(1)}&=\frac{1}{\omega_{eff}}\sum_{n\neq 0}\frac{1}{2in}[\tilde{\mathcal{L}}^{(-n)},\tilde{\mathcal{L}}^{(n)}],\\
\mathcal{L}_{eff(2)}&=\frac{1}{\omega_{eff}^{2}}\{\sum_{n\neq 0}\frac{1}{2n^{2}}[\tilde{\mathcal{L}}^{(n)},\partial_{t}\tilde{\mathcal{L}}^{(-n)}]\\
&+\sum_{n\neq 0}\frac{1}{6n^{2}}[\tilde{\mathcal{L}}^{(n)},[\tilde{\mathcal{L}}^{(-n)},\tilde{\mathcal{L}}^{(0)}]]\\
&-\sum_{m,n\neq 0}\frac{1}{3mn}[\tilde{\mathcal{L}}^{(m)},[\tilde{\mathcal{L}}^{(n)},\tilde{\mathcal{L}}^{(-m-n)}]]\}.
\end{aligned}
\end{equation}

In contrast with the earlier studies that the generator $\mathcal{L}(t)$ is periodic in time with fixed frequency, the $\omega$ here is replaced by the instantaneous frequency $\omega_{eff}$, all components $\tilde{\mathcal{L}}^{(n)}[\mathbf{p}(t)]$ of $\tilde{\mathcal{L}}$ can have slow time dependence and a non-trivial term $[\tilde{\mathcal{L}}^{(n)},\partial_{t}\tilde{\mathcal{L}}^{(-n)}]$ appears in the second order term. Thus $\mathcal{L}_{eff}$ will acquire slow time dependence and generally not commute at different time. The approximate effective generator depends not only on the instantaneous values of parameters but also on the  rate of changes. As for the expansions of $\Omega(t)$ the derivative of $\omega_{eff}$ also appears in the second order term,
\begin{equation}\label{oexp}
\begin{aligned}
\Omega_{(1)}^{(n)}&=\frac{-i}{\omega_{eff}n}\tilde{\mathcal{L}}^{(n)},\\
\Omega_{(2)}^{(n)}&=\frac{-1}{\omega_{eff}^2}\{\frac{1}{2n^{2}}[\tilde{\mathcal{L}}^{(0)},\tilde{\mathcal{L}}^{(n)}]\\
&+\sum_{m\neq 0}\frac{1}{2mn}[\tilde{\mathcal{L}}^{(n-m)},\tilde{\mathcal{L}}^{(m)}]\\
&-\frac{1}{n^{2}}[\partial_{t}\tilde{\mathcal{L}}^{(n)}-(\dot{\omega}_{eff}/\omega_{eff})\tilde{\mathcal{L}}^{(n)}]\}.
\end{aligned}
\end{equation}
These equations together give  the effective generator with slow time dependence $\mathcal{L}_{eff}=\mathcal{L}_{eff(0)}+\mathcal{L}_{eff(1)}+\mathcal{L}_{eff(2)}+\mathcal{O}(\omega_{eff}^{-3})$ and the corresponding micro-motion operator $\mathcal{D}(\theta+\omega t,t)=exp[\Omega_{(1)}(\theta+\omega t,t)+\Omega_{(2)}(\theta+\omega t,t)+\mathcal{O}(\omega_{eff}^{-3})]$ where $\Omega_{(m)}(\theta+\omega t,t)=\sum_{n\neq0}\Omega_{(m)}^{(n)}(t)e^{in(\theta+\omega t)}$ (We use the relation $\sum_{n=-\infty}^{\infty}\mathcal{D}^{(n)}(t)e^{in(\theta+\omega t)}=exp[\sum_{n=-\infty}^{\infty}\Omega^{(n)}(t)e^{in(\theta+\omega t)}]$ when we calculate the summation).

Discussions on the  present expansions  are in order. When $\omega_{eff}\rightarrow \infty$, $\mathcal{L}_{eff}\rightarrow\mathcal{L}_{eff(0)}$, i.e., the generator takes  the zeroth  Fourier component, while the micro-motion $\mathcal{D}(\theta+\omega t,t)\rightarrow\mathcal{I}$ approaches identity in this case.  When   the frequency $\omega$ and slow-varying parameters $\mathbf{p}$ are independent of time, the expansions Eq.(\ref{leff expansion}) and Eq.(\ref{oexp}) reduce to the time-independent form that are the same as that in  previous works \cite{bukov2015,eckardt2015,goldman2014,haddadfarshi2015},  where the dynamics is strictly  periodic.

\section{Examples}
In this section we illustrate our theory with two examples. The first example consists of  a $spin\textendash\frac{1}{2}$ particle coupled to a time-dependent magnetic field, and the second one is a harmonic oscillator driven by a time-dependent field. Both of them are subject to decoherence.

The generators that describe the two examples have a similar form
\begin{equation} \label{mm1}
\begin{aligned}
\mathcal{L}(t)&(\circ)=\tilde{\mathcal{L}}[\theta+\omega t,\mathbf{p}(t)](\circ)\\
=-i[H(t),\circ]&+\gamma(2X_{-}\circ X_{+}-\{X_{+}X_{-},\circ\}),
\end{aligned}
\end{equation}
where $X_{+}=X_{-}^{\dagger}$ represents system-bath interaction and $\gamma$ is decay rate (system-bath
coupling strength), the Hamiltonian $H(t)=H^{(0)}[\mathbf{p}(t)]+H^{(-1)}[\mathbf{p}(t)]e^{-i(\theta+\omega t)}+H^{(1)}[\mathbf{p}(t)]e^{i(\theta+\omega t)}$ with $H^{(n)}[\mathbf{p}(t)]$ ($n=-1,0,1$) depends on the slowly-varying parameters $\mathbf{p}(t)$. Then expansion Eq.(\ref{leff expansion}) reduces to (leave out trivial term $\mathcal{L}_{eff(0)}$)
\begin{equation}\label{leff simple}
\begin{aligned}
\mathcal{L}_{eff(1)}(\circ)=&\frac{1}{i\omega_{eff}}[[H^{(1)},H^{(-1)}],\circ],\\
\mathcal{L}_{eff(2)}(\circ)=&\frac{1}{2\omega_{eff}^{2}}\sum_{n=-1,1}\{[[\partial_{t}H^{(-n)},H^{(n)}],\circ]\\
+&i[[H^{(n)},[H^{(-n)},H^{(0)}]],\circ]\\
+&\gamma[\{[H^{(n)},[H^{(-n)},X_{+}X_{-}]],\circ\}\\
+&2[H^{(n)},X_{-}]\circ[X_{+},H^{(-n)}]\\
+&2[H^{(-n)},X_{-}]\circ[X_{+},H^{(n)}]\\
+&2[H^{(n)},[X_{-},H^{(-n)}]]\circ X_{+}\\
+&2 X_{-}\circ[H^{(n)},[X_{+},H^{(-n)}]]]\}.
\end{aligned}
\end{equation}
As shown, at a long time scale,  both the effective Hamiltonian and system-bath interaction are modified by driving field characterized by $H^{(-1)}[\mathbf{p}(t)]=H^{(1)}[\mathbf{p}(t)]^{\dagger}$ and frequency $\omega$.

\subsection{Spin\textendash$\frac{1}{2}$ particle}
For a $spin\textendash\frac{1}{2}$ particle in a fast  oscillating magnetic field with additional slow modulation, the system can be described by the generator in Eq.(\ref{mm1}) with $X_{-}=\sigma_{-}$, $X_{+}=\sigma_{+}$ and $H(t)=\alpha\mathbf{B}_{total}(t) \cdot \boldsymbol{\sigma}$, where $\alpha$ is the coupling constant and magnetic field $\mathbf{B}_{total}(t)=\mathbf{B}_{0}(t)+\mathbf{B}(t)e^{i(\theta+\omega t)}+\mathbf{B}^{*}(t)e^{-i(\theta+\omega t)}$. Here the slowly-varying parameter $\mathbf{p}(t)=\{\mathbf{B}_{0}(t),\mathbf{B}(t)\}$. $\omega$ is the angular frequency of the fast oscillating. Using Eq.(\ref{leff simple}), we get
\begin{equation}\label{sl}
\begin{aligned}
\mathcal{L}_{eff(1)}(\circ)=&[\frac{2 \alpha^{2}}{\omega_{eff}}[\mathbf{B}\times\mathbf{B}^{*}]\cdot \boldsymbol{\sigma},\circ],\\
\mathcal{L}_{eff(2)}(\circ)=&\frac{\alpha^{2}}{2\omega_{eff}^{2}}\{[-4i \alpha[\mathbf{B}\times[\mathbf{B}^{*}\times \mathbf{B}_{0}]]\cdot \boldsymbol{\sigma},\circ]\\
+&[2i[\partial_{t}\mathbf{B}^{*}\times\mathbf{B}]\cdot \boldsymbol{\sigma},\circ]\\
-&4\gamma[i\{[\mathbf{B}\times[\mathbf{B}^{*}\times[\mathbf{e}_{+}\times\mathbf{e}_{-}]]]\cdot \boldsymbol{\sigma},\circ\}\\
+&4[\mathbf{B}\times\mathbf{e}_{-}]\cdot\boldsymbol{\sigma}\circ[\mathbf{e}_{+}\times\mathbf{B}^{*}]\cdot\boldsymbol{\sigma}\\
+&2[\mathbf{B}\times[\mathbf{e}_{-}\times\mathbf{B}^{*}]]\cdot\boldsymbol{\sigma}\circ\sigma_{+}\\
+&2\sigma_{-}\circ[\mathbf{B}\times[\mathbf{e}_{+}\times\mathbf{B}^{*}]]\cdot\boldsymbol{\sigma}]\\
+&(\mathbf{B}\leftrightarrow\mathbf{B}^{*})\},
\end{aligned}
\end{equation}
where $\mathbf{e}_{\pm}=(\mathbf{e}_{x}\pm i\mathbf{e}_{y})/2$, $\mathbf{e}_{x}$ and $\mathbf{e}_{y}$ are the unit vectors along the $x\textendash$ and $y\textendash$ direction, respectively.

Consider the case that  $\mathbf{B}_{total}(t)$ rotates rapidly around $\mathbf{e}_{x}$ in the $y\textendash z$ plane with slowly-varying angular velocity, and  denote the included angle between $\mathbf{B}_{total}$ and $\mathbf{e}_{y}$ as $\theta+\omega t$. We have $\mathbf{B}_{total}(t)=[0,\cos(\theta+\omega t),\sin(\theta+\omega t)]$ (we will set $\theta=0$ hereafter for concrete), i.e., $\mathbf{B}_{0}=0$, $\mathbf{B}=\frac{1}{2}[0,1,-i]$ and the angular velocity $\omega_{eff}=\partial_{t}(\omega t)$ changes slowly. Substituting  these into Eq.(\ref{sl}), we get
\begin{displaymath}
\mathcal{L}_{eff(1)}(\circ)=i(\alpha^{2}/\omega_{eff})[\sigma_{x},\circ].
\end{displaymath}
Because the first two terms of $\mathcal{L}_{eff(2)}$ vanish, the third term of $\mathcal{L}_{eff(2)}$ proportional to $\omega_{eff}^{-2}$ is also negligible when $\omega_{eff}$ is relatively large, then $\mathcal{L}_{eff}\approx\mathcal{L}_{eff(0)}+\mathcal{L}_{eff(1)}$ is a good approximation.

When the effective generator $\mathcal{L}_{eff}$ changes slowly enough, the system is expected to follow the instantaneous steady state of  $\mathcal{L}_{eff}$ except the micro-motion. The instantaneous steady state of  $\mathcal{L}_{eff}$ up to the first order  in $\omega_{eff}^{-1}$ is
\begin{displaymath}
\rho_{s}=1/2-[\gamma\omega_{eff}\alpha^{2}\sigma_{y}+\frac{(\gamma\omega_{eff})^2}{2}\sigma_{z}]/[2\alpha^{4}+(\gamma\omega_{eff})^2].
\end{displaymath}
We can see that the instantaneous steady state $\rho_{s}$ depends on the coupling constant $\alpha$ and the product of bath coupling strength $\gamma$ and effective frequency $\omega_{eff}$. When $\gamma\omega_{eff}\gg\alpha^2$ or $\gamma\omega_{eff}\ll\alpha^2$, $\rho_{s}\rightarrow|\downarrow\rangle\langle\downarrow|$ or $(|\downarrow\rangle\langle\downarrow|+|\uparrow\rangle\langle\uparrow|)/2$ respectively. The steady behavior can be manipulated by the driving frequency or the system-bath coupling $\gamma$.

\begin{figure}
	\includegraphics[scale=0.62]{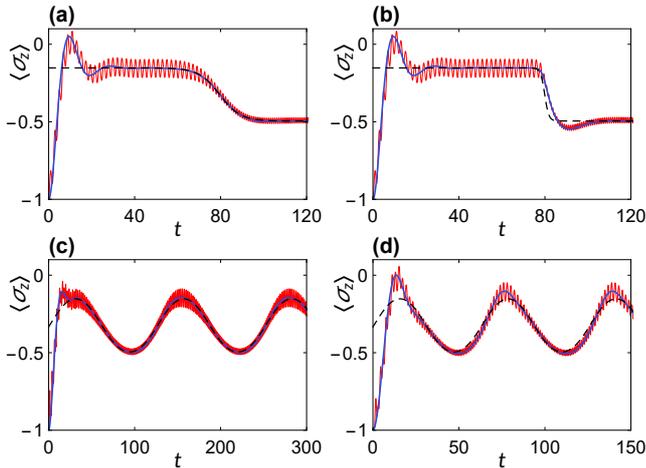}
	\caption{The expectation value of $\sigma_{z}$ (dimensionless) versus time. Parameters chosen are $\alpha=1/2$, $\gamma=0.1$ for (a), (b), (c) and (d). In the first two plots $\omega_{eff}$ increases slowly from $\omega_{i}$ to $\omega_{f}$ by $\omega_{eff}=\frac{\omega_{f}+\omega_{i}}{2}+\frac{\omega_{f}-\omega_{i}}{2}\tanh[(t-t_{0})/\tau_{a,b}]$ and in the last two plots $\omega_{eff}$ changes periodically from $\omega_{i}$ to $\omega_{f}$, $\omega_{eff}=\frac{\omega_{f}+\omega_{i}}{2}+\frac{\omega_{f}-\omega_{i}}{2}\sin(t/\tau_{c,d})$, where $\omega_{i}=1.5$, $\omega_{f}=3.5$, $t_{0}=8/\gamma$, $\tau_{a}=1/\gamma$, $\tau_{b}=1/5\gamma$, $\tau_{c}=2/\gamma$, $\tau_{d}=1/\gamma$. Red thin line is the numerical result given by $\mathcal{L}$. Blue thick line is obtained by  $\mathcal{L}_{eff}$. Black dashed line is the result by $\rho_{s}$.}\label{varyfloquet1}
\end{figure}

Fig.\ref{varyfloquet1} shows the difference between the average  of $\sigma_{z}$ in states given by $\mathcal{L}_{eff(0)}+\mathcal{L}_{eff(1)}$, $\mathcal{L}$ and $\rho_{s}$, respectively. Here we consider    $\omega_{eff}$ changing  in two different ways---In the first case, $\omega_{eff}$ increases from $\omega_{i}$ to $\omega_{f}$, once the maximum has been reached, $\omega_{eff}$ remains constant. In another case, $\omega_{eff}$ changes periodically.

In Fig.\ref{varyfloquet1} (a) and (c), $\omega_{eff}$ changes slowly enough compared with $1/\gamma$, the results obtained by instantaneous steady state $\rho_{s}$ are almost the same as that by $\mathcal{L}_{eff}$. In Fig.1 (b) and (d), $\omega_{eff}$ changes a little faster that introduce a small departure between the results obtained by $\rho_{s}$ and $\mathcal{L}_{eff}$.

\begin{figure}
	\includegraphics[scale=0.51]{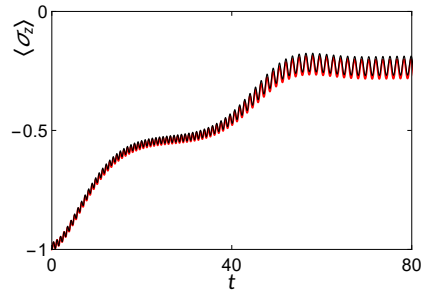}
	\caption{The expectation value of $\sigma_{z}$ (dimensionless) versus time. $\omega_{eff}$ has the same form as in Fig.\ref{varyfloquet1} (a), the parameters chosen are $\alpha=1/2$, $\gamma=0.1$, $\omega_{i}=4$, $\omega_{f}=2$, $t_{0}=4/\gamma$, $\tau=1/\gamma$. Red thick line is the numerical result by $\mathcal{L}$, black thin line is given by combining  approximate micro-motion operator $\mathcal{D}_{(1)}$ and $\mathcal{L}_{eff}$.}\label{varyfloquet2}
\end{figure}

The results by $\mathcal{L}$ and $\mathcal{L}_{eff}$ are fairly consistent except for the fast oscillation due to the micro-motion.  To the first order in $\omega_{eff}^{-1}$, the operator $\Omega$ can be calculated by Eq.(\ref{oexp}),
\begin{displaymath}
\Omega_{(1)}(\circ)=-i(\alpha/\omega_{eff})[\sigma_{y}\sin{\omega t}-\sigma_{z}\cos{\omega t},\circ],
\end{displaymath}
and the micro-motion operator by $\mathcal{D}(\circ)=e^{\Omega_{(1)}(\circ)+\mathcal{O}(\omega_{eff}^{-2})}$. Up to the first order of $\omega_{eff}^{-1}$, $\mathcal{D}(\circ)$ goes back to its initial value after $\omega t$ changing $2\pi$ with possible correction caused by slowly-varying of $\omega_{eff}$. As show in Fig.\ref{varyfloquet1} (a) red thin line, when $\omega_{eff}$ increase, the amplitude of oscillation decrease gradually and the period of oscillation approximately equals to $2\pi/\omega_{eff}$. Fig.\ref{varyfloquet2} shows the results obtained by the combination of  $\mathcal{D}_{(1)}(\circ)=e^{\Omega_{(1)}(\circ)}$ and  $\mathcal{L}_{eff}$. A comparison with the exact result is also carried out. It is clear that the first order micro-motion together with the effective Lindblad  $\mathcal{L}_{eff}$ can give a fairly accurate time evolution for  the system.
It is worth addressing that the higher order terms of $\Omega$  contain  dissipative  effect due to the system-bath interactions, though in the first order approximation the micro-motion  governed by $\Omega$ is well approximate by a unitary evolution $\mathcal{D}_{(1)}(\circ)$.

The situation would  be quite different when we consider periodically  modulated  system-bath interaction, the major contribution of the micro-motion is dissipative  \cite{kamleitner2011,dai2016} that can also be calculated by Eq.(\ref{oexp}).

\begin{figure}
	\includegraphics[scale=0.63]{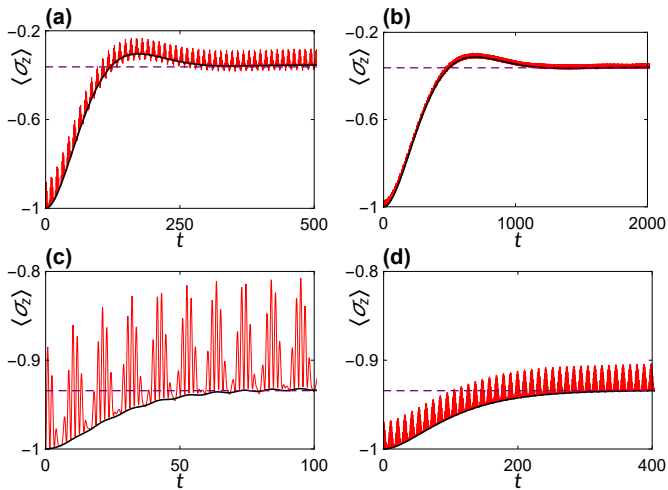}
	\caption{The expectation value of $\sigma_{z}$ (dimensionless) versus time. The parameters chosen are $\omega_{c}=0.3$, $\alpha=1/2$ for all plots.
	In (a) and (c) $\omega=2$, in (b) and (d) $\omega=4$. $\gamma=1\times10^{-2},2.5\times10^{-3},5\times10^{-2},1.25\times10^{-2}$ for (a), (b), (c) and (d) respectively. Red thin line is the numerical result by $\mathcal{L}$, black thick line obtained by approximate $\mathcal{L}_{eff}$ up to the second order in $\omega_{eff}^{-1}$, dashed purple line indicates $Tr[\sigma_{z}\rho_{\infty}]$. }\label{varyfloquet3}
\end{figure}

To demonstrate the effect of the  non-trivial term $\sim \partial_{t}\mathbf{B}^{*}\times\mathbf{B}$ that depends on the change rate of slow-varying parameters $\mathbf{p}(t)=\{\mathbf{B}_{0}(t),\mathbf{B}(t)\}$. We may set  $\mathbf{B}_{0}=0$, $\mathbf{B}=\frac{1}{2}[0,\cos{\omega_{c}t},\sin{\omega_{c}t}]$ as an example. Such a specific choice corresponds a magnetic field $\mathbf{B}_{total}(t)=\cos(\theta+\omega t)[0,\cos{\omega_{c}t},\sin{\omega_{c}t}]$ that rotate slowly around $\mathbf{e}_{x}$ in the $y\textendash z$ plane with constant angular velocity $\omega_{c}$ and the strength of magnetic field changes quickly with fixed angular frequency $\omega$. By Eq.(\ref{sl}), the first order term of $\mathcal{L}_{eff}$ vanishes and the second order term is
\begin{displaymath}
\mathcal{L}_{eff(2)}(\circ)=(\alpha^{2}/\omega^{2})\{-i[\frac{\omega_{c}}{2}\sigma_{x},\circ]-\gamma\sum_{i,j=0}^{3}C_{ij}\sigma_{i}\circ\sigma_{j}\},
\end{displaymath}
where the coefficient matrix takes,
\begin{displaymath}
C=32\cos^{2}{\omega_{c}t}
\begin{pmatrix}
0&0&\tan{\omega_{c}t}&-1&\\
0&2&i&i\tan{\omega_{c}t}&\\
\tan{\omega_{c}t}&-i&0&\tan{\omega_{c}t}&\\
-1&-i\tan{\omega_{c}t}&\tan{\omega_{c}t}&-2&
\end{pmatrix}.
\end{displaymath}
When the frequency $\omega$ is large, the explicit time dependence of $\mathcal{L}_{eff}$ caused by matrix $C$ can be neglected, because it represents a small correction to the decay  in the zeroth order term of $\mathcal{L}_{eff}$. The major contribution is the Hamiltonian part in $\mathcal{L}_{eff(2)}$. In this case the steady state reads,
\begin{displaymath}
\rho_{\infty}=1/2+(\alpha^{2}\omega_{c}\gamma\omega^{2}\sigma_{y}
-\gamma^{2}\omega^{4}\sigma_{z})/(2\gamma^{2}\omega^{4}+\alpha^{4}\omega_{c}^{2}).
\end{displaymath}
As show in Fig.\ref{varyfloquet3}, the average of $\sigma_{z}$ obtained by $\rho_{\infty}$ and $\mathcal{L}_{eff}$ is very close, and  $\rho_{\infty}$ depends on $\alpha^2 \omega_{c}$ and $\gamma \omega^2$. The change rate of slowly-varying parameters characterized by $\omega_{c}$ can also affect  the long time dynamics.

The operator $\Omega$ up to the first order in $\omega_{eff}^{-1}$ can be written as
\begin{displaymath}
\Omega_{(1)}(\circ)=-2i(\alpha\sin{\omega t}/\omega)[\mathbf{B}\cdot\boldsymbol{\sigma},\circ],
\end{displaymath}
and the micro-motion operator takes $\mathcal{D}(\circ)=e^{\Omega_{(1)}(\circ)+\mathcal{O}(\omega_{eff}^{-2})}$. For fixed $\mathbf{B}$,  $\mathcal{D}_{(1)}(\circ)=e^{\Omega_{(1)}(\circ)}$ is periodic in time with period $2\pi/\omega$. The slow variation of $\mathbf{B}$ introduces additional time dependence to the dynamics. The approximate micro-motion operator becomes identity map when $t=n\pi/\omega$ with $n$ positive integers. As shown in Fig.\ref{varyfloquet3} (c), the red thin line touches the black thick line when $t=n\pi/\omega$.

\subsection{Harmonic oscillator}
For a driven harmonic oscillator coupled to a damped environment, $\mathcal{L}(t)(\circ)=-i[H(t),\circ]+\gamma(2a\circ a^{\dagger}-\{a^{\dagger}a,\circ\})$ with $H(t)=\omega_{0} a^{\dagger}a+f(t)\cos{\omega_{d}t}(a+a^{\dagger})$. Here $f(t)$ and $\omega_{d}$ are the amplitude and frequency of driven field, respectively. We consider the situation where $\omega_{0}\approx\omega_{d}$.  Transforming the master equation  to the interaction picture, i.e.,  $\rho_{I}(t)=e^{i\omega_{0} a^{\dagger}at}\rho(t)e^{-i\omega_{0} a^{\dagger}at}$, we have
\begin{displaymath}
\partial_{t}\rho_{I}=-i[H_{I}(t),\rho_{I}]+\gamma(2a\rho_{I}a^{\dagger}-a^{\dagger}a\rho_{I}-\rho_{I} a^{\dagger}a),
\end{displaymath}
where $H_{I}(t)=f(t)\cos{\omega_{d}t(ae^{-i\omega_{0} t}+a^{\dagger}e^{i\omega_{0} t})}$. Assume $\omega_{0}\approx\omega_{d}$, then $H_{I}(t)=H_{I}^{(0)}+H_{I}^{(1)}e^{i\omega t}+H_{I}^{(-1)}e^{-i\omega t}$ depends on slowly-varying parameter $\mathbf{p}(t)=\{f(t),e^{i(\omega_{d}-\omega_{0})t}\}$,  and $\omega$ in this case $\omega=\omega_{0}+\omega_{d}$. Here, $H_{I}^{(0)}=\frac{1}{2}f(t)(ae^{i(\omega_{d}-\omega_{0})t}+h.c.)$, $H_{I}^{(1)}=\frac{1}{2}f(t)a^{\dagger}$, $H_{I}^{(-1)}=\frac{1}{2}f(t)a$.  Substituting  these equations into Eq.(\ref{leff simple}) and setting  $X_{-}=a$, $X_{+}=a^{\dagger}$, one can find that the first order and second order terms of $\mathcal{L}_{eff}$ vanish. So up to the second order in $\omega_{eff}^{-1},$ we have
\begin{displaymath}
\mathcal{L}_{eff}(\circ)\approx-\frac{i}{2}f(t)[ae^{i(\omega_{d}-\omega_{0})t}+h.c.,\circ]+\gamma[2a\circ a^{\dagger}-\{a^{\dagger}a,\circ\}],
\end{displaymath}
and up  to the first order in $\omega_{eff}^{-1}$, we have
\begin{displaymath}
\Omega_{(1)}(\circ)=-(f(t)/2\omega_{eff})[a^{\dagger}e^{i\omega t}-ae^{-i\omega t},\circ].
\end{displaymath}
So $\mathcal{D}_{(1)}(\circ)=e^{\Omega_{(1)}(\circ)}=e^{\alpha a^{\dagger}-\alpha^{*}a}(\circ)e^{\alpha^{*}a-\alpha a^{\dagger}}$ is just the displacement operator with parameter $\alpha=-f(t)e^{i\omega t}/2\omega_{eff}$.

\begin{figure}
	\includegraphics[scale=0.6]{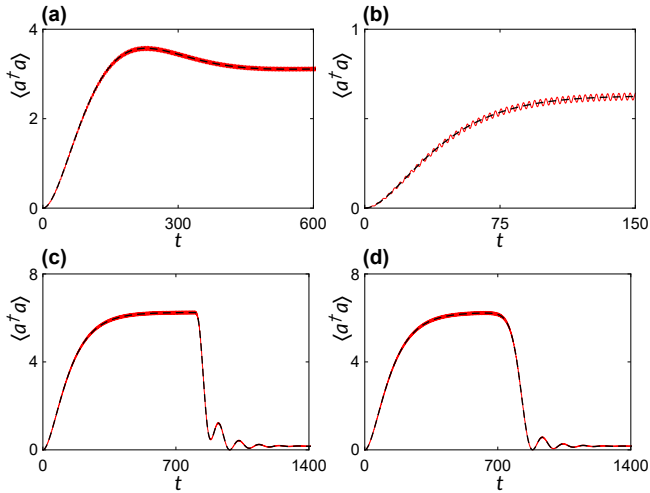}
	\caption{The expectation value of $a^{\dagger}a$ (dimensionless) versus time. We choose  frequency $\omega_{0}=1$ and amplitude of the  driven field $f(t)=0.05$ in all plots, decay rate $\gamma=10^{-2},3\times10^{-2},10^{-2},10^{-2}$ for (a), (b), (c) and (d), respectively. In the first two plots, the driven frequency are fixed to be $\omega_{d}=1.01$, while in the last two plots, it changes via $\partial_{t}(\omega_{d}t)=\frac{\omega_{f}+\omega_{i}}{2}
	+\frac{\omega_{f}-\omega_{i}}{2}\tanh[(t-t_{0})/\tau_{c,d}]$. Where $\omega_{i}=1$, $\omega_{f}=1.06$, $t_{0}=8/\gamma$. $\tau_{c}=1/20\gamma$ and $\tau_{d}=1/\gamma$ for figure (c) and (d), respectively. Red thin line is the numerical result given by $\mathcal{L}$, black dashed line was obtained by  $\mathcal{L}_{eff}$ up to the second order in $\omega_{eff}^{-1}$.}\label{varyfloquet4}
\end{figure}

We plot  Fig.\ref{varyfloquet4} (a) and (b) for fixed $\omega_{d}$ and $f(t)$, and from the figures we find that though   $\mathcal{L}_{eff}$ is time-dependent with period $|2\pi/(\omega_{d}-\omega_{0})|$, the average  number   $\langle a^{\dagger}a\rangle$ obtained by $\mathcal{L}_{eff}$ reaches a steady value  due to the asymptotic behavior of $\rho_{eff}(t)$, $\rho_{eff}(t)\rightarrow e^{-i(\omega_{d}-\omega_{0})a^{\dagger}at}\rho_{c}(\infty)e^{i(\omega_{d}-\omega_{0})a^{\dagger}at}$ with   $\rho_{c}$ governed by,
\begin{displaymath}
\mathcal{L}_{c}(\circ)=i[(\omega_{d}-\omega_{0})a^{\dagger}a-\frac{f}{2}(a+a^{\dagger}),\circ]+\gamma[2a\circ a^{\dagger}-\{a^{\dagger}a,\circ\}].
\end{displaymath}
Because $\rho_{I}(t)=\mathcal{D}\rho_{eff}(t)\approx\mathcal{D}_{(1)}\rho_{eff}(t)$, we have $Tr[\rho_{I}(t)a^{\dagger}a]\rightarrow Tr[\rho_{c}(\infty)a^{\dagger}a]-2|\alpha|\eta\cos(2\omega_{d}t+\delta)+|\alpha|^{2}$ with $\eta e^{i \delta}=Tr[\rho_{c}(\infty)a^{\dagger}]$. The result is  shown in Fig.\ref{varyfloquet4} (b), see the red thin line that oscillates with period $\pi/\omega_{d}$.

For varying effective frequency, e.g., $\omega_{eff}=\omega_{0}+\partial_{t}(\omega_{d}t)$ in Fig.\ref{varyfloquet4} (c) and (d), the frequency $\omega_{d}$ slowly changes  from the resonant point $\omega_{0}=\omega_{d}$.  The results obtained by effective generator $\mathcal{L}_{eff}$ is also consistent  with the exact dynamics except the small oscillation given by the  micro-motion operator.

\section{Conclusion and discussions}
In this paper, we have extended  the open-system  Floquet theorem  to a more general situation. The extended formulism permits us to include slow-varying parameters that break down the periodicity considered in the earlier open-system Floquet theorem. This extension has been done by removing  the fast periodic  term from  the time-evolution operator(or generator) to obtain an effective generator that depends on   time slightly. The slow-varying generator leads to an  asymptotic solution combining  with the micro-motion operator. We also give a hight-frequency expansion to  the effective generator and micro-motion operator,  showing  that the first two orders of the expansions agree well with the exact dynamics.

Compared with the conventional Floquet formalism, the slow-varying parameter can play an important role to control the long time dynamics. A natural extension of our result is to consider a system with two periodically drivings. One is small while another is very large. In terms of frequencies, this case  is,  $\omega_{1}\gg\omega_{2}$, our results can be applied easily to this situation.

\textbf{Finally, we would like to point out that the formulism presented in this paper is limited to weak system-bath couplings. As we use the master equation as the starting points of discussion.}

\section*{ACKNOWLEDGMENTS}
This work is supported by the National Natural Science Foundation of China (Grant Nos. 11534002, 61475033).

\end{document}